\documentclass[reprint,
superscriptaddress,
nofootinbib,
nobibnotes,
 amsmath,amssymb,
 aps,
pre,
]{revtex4-1}

\usepackage{tabularx}
\usepackage{import}
\usepackage{multirow}
\usepackage[table,xcdraw]{xcolor}
\usepackage{amsfonts}
\usepackage{amssymb}
\usepackage{graphicx}
\usepackage[colorlinks=true,citecolor=green,linkcolor=red,urlcolor=blue]{hyperref}
\usepackage{mathrsfs}
\usepackage{array}
\usepackage[fleqn,tbtags]{mathtools}
\usepackage{physics}
\usepackage{tabularx}
\usepackage[table]{xcolor}
\usepackage{bbold}
\usepackage{ucs}
\usepackage{xcolor}

\newcommand{\ul}[1]{\underline{#1}}

\newcommand{\rom}[1]{\uppercase\expandafter{\romannumeral #1\relax}}

\begin{document}

\title{Crisis Propagation in a Heterogeneous Self-Reflexive DSGE Model}

\author{Federico Guglielmo Morelli}
\affiliation{LPTMC, UMR CNRS 7600, Sorbonne Universit\'e, 75005 Paris, France}
\affiliation{LadHyX, UMR CNRS 7646, Ecole Polytechnique, 91128 Palaiseau, France}
\affiliation{Chair of Econophysics \& Complex Systems, Ecole polytechnique, 91128 Palaiseau, France}

\author{Michael Benzaquen} \email{michael.benzaquen@polytechnique.edu}
\affiliation{LadHyX, UMR CNRS 7646, Ecole Polytechnique, 91128 Palaiseau, France}
\affiliation{Chair of Econophysics \& Complex Systems, Ecole polytechnique, 91128 Palaiseau, France}
\affiliation{Capital Fund Management, 23-25, Rue de l'Universit\'e 75007 Paris, France}

\author{Jean-Philippe Bouchaud}
%\affiliation{LPTMC, UMR CNRS 7600, Sorbonne Universit\'e, 75252 Paris Cedex 05, France}
\affiliation{Chair of Econophysics \& Complex Systems, Ecole polytechnique, 91128 Palaiseau, France}
\affiliation{Capital Fund Management, 23-25, Rue de l'Universit\'e 75007 Paris, France}
\affiliation{Acad\'emie des Sciences, Quai de Conti, 75006 Paris, France}
\author{Marco Tarzia}
\affiliation{LPTMC, UMR CNRS 7600, Sorbonne Universit\'e, 75005 Paris, France}
\affiliation{Institut  Universitaire  de  France,  1  rue  Descartes,  75005  Paris,  France\medskip}

% Please give the surname of the lead author for the running footer

\date{\today}

\date{\today}

\begin{abstract}
We study a self-reflexive DSGE model with heterogeneous households, aimed at characterising the impact of economic recessions on the different strata of the society. Our framework allows to analyse the combined effect of income inequalities and confidence feedback mediated by heterogeneous social networks. By varying the parameters of the model, we find different crisis typologies: loss of confidence may propagate mostly within high income households, or mostly within low income households, with a rather sharp crossover between the two. We find that crises are more severe for segregated networks (where confidence feedback is essentially mediated between agents of the same social class), for which cascading contagion effects are stronger. 
%For non seregated networks, inter-social class interactions tend to dampen the propagation of pessimism, because agents belonging to different social classes have different sensitivities to economic shocks.
For the same reason, larger income inequalities tend to reduce, in our model, the probability of global crises.
%We also find that large income inequalities decrease the  probability of global crises. 
Finally, we are able to reproduce a perhaps counter-intuitive empirical finding: in countries with higher Gini coefficients, the consumption of the lowest income households tends to drop less than that of the highest incomes in crisis times. 
%Confidence feedback is mediated through the social network of each agent, which we assume to be either within social classes only (segregated network), or across social classes (non segregated network).
% we find, perhaps counter-intuitively, that in more unequal countries (with high Gini coefficients), the consumption of the lowest income households tends to drop less than that of the highest incomes. %This trend is mostly driven by the Gini coefficient and not by the country GDP/capita.
%Our model can be calibrated to reproduce such an empirical finding -- in fact only a small region of the parameters is compatible with the sign of the empirical effect. In particular, we find that the segregated network hypothesis is strongly favoured by the data.         

%which differ by their income and consumption levels, and by their social network. 
\end{abstract}
\maketitle

%% Introduction %%

%% Overview
\section{Introduction}

 Dynamic Stochastic General Equilibrium (DSGE)  models \cite{gali2015monetary,sbordone2010policy} still constitute the workhorse for monetary policy around the world \cite{clarida2002simple,blanchard2007real}. Yet, their poor performance during the 2008 global financial crisis (GFC)
\cite{trichet2010reflections,blanchard2014danger,stiglitz2018modern,christiano2018dsge} 
 have raised a number of questions about their predictive power \cite{rebuilding_macro}. In recent years, efforts have been made to include in these models ingredients that were sorely missing from the benchmark model \cite{gali2018state}, like financial markets. Attempts have been made to move away from the ``Representative Agent'' paradigm, by including different categories of households -- hand-to-mouth vs.~well-off in TANK (Two-Agent New Keynesian) models \cite{mankiw2000savers, bilbiie2008accounts, bilbiie2017new,debortoli2017monetary} -- or heterogeneous households with a continuum of possible accumulated wealth, as in HANK (Heterogeneous Agent New Keynesian) models \cite{oh2012targeted,kaplan2018monetary, ragot2018heterogeneous,ravn2016macroeconomic,gornemann2016doves,bilbiie2018monetary}; see also 
 \cite{Farmer2019behaviour} for a different approach leading to emergent heterogeneities. 
 
 Parallel to these developments, macroeconomic Agent Based Models (ABM) are slowly gaining traction \cite{dawid, geanakoplos2012getting, gualdi2015tipping, haldane2018interdisciplinary}. The versatility of ABM allows one to investigate the role of interactions and heterogeneities that often lead to interesting (and sometimes surprising) effects at the aggregate level, as a consequence of non-linearities and feedback loops that are absent from classical economic models (see e.g.~\cite{gualdi2015tipping}). Still, ABM are regarded with suspicion by many macroeconomic luminaries, who prefer to stick with ``microfounded'' models where agents solve an inter-temporal optimisation problem with a budget constraint \cite{blanchard2018future}. 
 
 A natural question is whether it is possible to extend DSGE models in a direction that would bridge the gap with ABM, in particular by including social interactions and heterogeneities that are the strong selling points of ABM. Our starting point is our recent work \cite{guglielmo2019confidence} where we investigated a multi-household DSGE model in which past aggregate consumption impacts the confidence, and therefore consumption propensity, of individual households. We found that our minimal setup was already extremely rich, leading to a variety of realistic output dynamics, in particular the appearance of crises where consumption drops as a result of an initial exogenous shock, amplified by a collapse of confidence. But while we modelled interactions and feedback loops in \cite{guglielmo2019confidence}, we did not account for possible income heterogeneities and network effects. 
 
 Here we build upon such ideas and introduce agents that can be assigned different characteristics, such as skill and social environment. At each time step, the consumption level of each household %Each household, with a certain income, 
 is partially determined by % its consumption level partly based on 
 the past realised consumption of its neighbouring agents in a network of social interactions. As we shall see, we find that the phenomenon that we discussed in our previous paper \cite{guglielmo2019confidence}, namely the appearance of endogenous demand driven crises, is now more complex, with much more structured consumption crashes that are either restricted to some stratum of the population, or affect the whole population, depending on the distribution of wages and the structure of the social network. 

  Although we stick with the basic tenets of standard macroeconomic models, our approach departs in several ways from the %we depart from the 
 recent heterogeneous extensions of the DSGE models mentioned above (e.g. TANK or HANK models). % in several ways. 
 First, our heterogeneities affect both income and the structure of the %positions in the social 
 interaction network, the latter being absent (and irrelevant) in TANK/HANK approaches.  %, that are different for all agents in the model. 
 Second, our heterogeneities are static (low skill workers do not become high skill workers, and the social network is ``frozen''), whereas in HANK models earnings are dynamical variables as agents self-insure against possible loss of wages in the future. In reality, a mixture of the two should be expected: both static and dynamically generated heterogeneities are likely to be present in the population. Third, social interactions lead to self-amplified confidence collapse, absent in TANK/HANK models whereas it is the main feature of our model. 

Our model allows one to explore a variety of different possible scenarios. In particular we highlight the presence of two transitions between different regimes of the dynamical evolution of the economy. Varying the parameters that describe how the confidence threshold and the sensitivity to economic fluctuations scale with wages, we find a rather sharp crossover between a regime in which crises affect mostly the wealthier part of the population to a regime in which the recessions involve mostly the agents with a low income level.

We also show that increasing the amplitude of exogenous shocks and/or increasing  the global confidence threshold, the model exhibits a transition from a regime in which global crises are extremely rare to a phase in which strong recessions involving the whole population occur with high frequency. We investigate the effect of the parameters that control the strength of the heterogeneities and the social (income-based) segregation, providing intuitive and transparent explanations of their impact on the economy. Our model is extremely versatile and is able to describe a variety of  realistic scenarios for crisis formation and its propagation across different social classes. 

%\textcolor{red}{As an illustration, in the end of the paper we show comparison with real data \ldots}

The paper is organised as follows. Section~\ref{sec:skill} is devoted to finding the analytical solution of our heterogeneous DSGE model, starting from the benchmark DSGE model as presented by Gali in~\cite{gali2015monetary}. 
In Sec.~\ref{sec:feedback} section we discuss the role and form of the feedback function, and we introduce wage heterogeneities. We define the network that  encodes interactions among agents.
In Sec.~\ref{sec:parameters} we discuss the choice of parameters, using our previous work~\cite{guglielmo2019confidence} as a reference. The qualitative discussion of the results is addressed in Sec.~\ref{sec:intuition}; we conduct an analysis of economic crises by studying both their distribution across the population and the frequency of the occurrence of global events. 
Finally, in Sec.~\ref{sec:data} we compare the numerical simulations of our model with real data, discussing potentials and limitations of our approach.

\section{Skill \& Wage Heterogeneities} \label{sec:skill}

As mentioned above, our starting point is our recent work~\cite{guglielmo2019confidence} where we presented a homogeneous version of a self-reflexive multi-agent DSGE model. The different ingredients of the model are summarized as follows.

\subsection{Households} 

We consider $N$ households $i=1, \cdots, N$ who maximise the discounted sum of their present and future utility:
%\textcolor{red}{be more precise. I would put the full expression of the stream of expected utility.}
\begin{eqnarray}
U_t^i &: =& f_t^i \log c_t^i - \gamma^i (n_t^i)^2 \,
\label{eq:utility}
\end{eqnarray}
where $c_t^i, n_t^i$  are the level of consumption and the amount of working hours of household $i$ at time $t$, $f_t^i$ the (possibly time dependent, see below) utility of consumption and $\gamma^i$ the disutility of work. Utility maximisation is subject to the classic budget constraint:
 \begin{eqnarray}
p_t c_t^i + \frac{B_t^i}{1+r_t} &=& n_t^i w_t^i + B_{t-1}^i \ ,
\label{eq:budget}
\end{eqnarray}
where $p_t$ the price level of goods, $w_t^i$ the wage of agent $i$ and $B_t^i$ the amount of bonds paying 1 at time $t+1$, the value of which being  $(1+r_t)^{-1}$ at time $t$, where $r_t$ is the interest rate. 

The maximisation is performed using Lagrange multipliers with respect to the quantities $c_t^i$, $n_t^i$, $B_t^i$. This gives the household state equation:
\begin{eqnarray}
c_t^i n_t^i &=& f_t^i \omega_t^i / \gamma^i\ ,  \quad  i=1, \dots, N,  
\label{eq:state}
\end{eqnarray}
where we introduced the real wages $\omega_t^i = w_t^i / p_t$. One also obtains the Euler equation governing inter-temporal substitution of consumption:
 \begin{eqnarray}
(c_{t}^i)^{-1}&=& (1+r_t)\beta\mathbb{E}_t \left[\frac{f^i_{t+1} (c_{t+1}^i)^{-1}}{1+\pi_{t+1}}\right] \label{eq:euler} \ ,
\end{eqnarray}
where $\pi_{t+1}=p_{t+1}/p_t - 1$ is the inflation rate. This equation will not be used in the rest of the paper, as we will not be concerned with inflation at this stage (for a short inroad into inflation within the homogeneous version of the model, see \cite{guglielmo2019confidence}).

\subsection{Firms} 
The production sector is made of a representative firm which uses different skills, corresponding to different productivity levels $z^i$ among agents. We posit that the firm level of production $Y_t$ is given by a Cobb-Douglas function with 
$\sum_i z^i n_t^i$ as the effective number of working hours: 
\begin{eqnarray}
    Y_t = \mathfrak{z}_t \frac{N^{\alpha}}{1-\alpha} {\left(\sum_i {z}^i n_t^i \right)^{1-\alpha}},
    \label{eq:production}
\end{eqnarray}
where $\mathfrak{z}_t$ is an overall productivity factor, subject to exogenous shocks, and $\alpha < 1$ is a parameter henceforth set to the standard value 1/3. The pre-factor $N^{\alpha}$ in~(\ref{eq:production}) ensures that both the aggregate consumption and the production are proportional to the size of the population $N$.  The firm's profit $\mathbb{P}_t$ is then given by:
\begin{equation}
    \label{eq:profit}
\frac{\mathbb{P}_t}{p_t} := \sum_i c_t^i - \sum_i \omega_t^i n_t^i,
\end{equation}
The firm maximises $\mathbb{P}_t$ with respect to the individual labour supply $n_t^i$, under the assumption that the market will clear, i.e. 
\begin{equation} \label{eq:marketclearing}
Y_t = \sum_i c_t^i \, .
\end{equation}
Such maximisation provides the following relation between real wage $\omega_t^i$ and  productivity $z^i$ of each agent:
\begin{eqnarray}
    {\omega_t^i}  = \mathfrak{z}_t \, \frac{z^i}{\mathcal{Z}_t^\alpha} , %\ \ \forall i \in [1, N] 
    \qquad \mathcal{Z}_t:= \frac1N \sum_j z^j n_t^j
    \label{eq:wage}
\end{eqnarray}
Using Eqs.~(\ref{eq:state}) and~(\ref{eq:wage}), the market clearing condition~(\ref{eq:marketclearing}) becomes: %\textcolor{blue}{(please check, I find an extra factor $(1-\alpha)$ and I have a $\gamma^i^2$ missing)}: 
%Letting $F_t^i = (f_t^i)^{1/2}/\gamma^i$ each solution needs to satisfy the relation \eqref{eq:rel}
\begin{equation} 
    \sum_i \frac{z^i}{n_t^i} \left[ \left(n_t^{i} \right)^2
     - \frac{f_t^i (1-\alpha)}{\gamma^i} \right] = 0 \ .
    \label{eq:rel}
\end{equation}
Given the set of $\{ \gamma^i \}$ and $\{ f_t^i \}$,  Eq. \eqref{eq:rel} describes an $N-1$~dimensional manifold where the solutions $n^i_t$ must lie. 

Now, plugging Eq.~\eqref{eq:wage} into the profit function~\eqref{eq:profit}, one finds that 
\[
\frac{\mathbb{P}_t}{N p_t} = \frac{\mathfrak{z}_t \alpha}{1-\alpha} \mathcal{Z}_t^{1-\alpha}.
\]
Thus, among the set of possible solutions described by \eqref{eq:rel} we retain the one maximising the sum $\mathcal{Z}_t$. Introducing again Lagrange multipliers, one can show that the optimal solution is given by 
\begin{equation}\label{def:F}
n_t^i = F_t^i :=\sqrt{(1 - \alpha) \frac{f_t^i}{ \gamma^i}}, \qquad \forall i,
\end{equation}
i.e. each term of the sum in Eq. \eqref{eq:rel} is zero. 
Combining Eqs.~\eqref{eq:wage} and \eqref{eq:state} finally yields:
\begin{equation}\label{eq:solution}
    c_t^i = \mathfrak{z}_t \, \frac{z^i F_t^i}{1-\alpha} \, \left(\frac1N \sum_j z^j F_t^j\right)^{-\alpha}  .
\end{equation}
Eqs.~\eqref{eq:wage} and \eqref{eq:solution} are our central theoretical results. Eq.~\eqref{eq:solution}, which appears to be new, gives the general solution of a generalised DSGE model with many heterogeneous agents, while keeping most of the original DSGE fully rational agent paradigm intact up to now. %Equation Eq. \eqref{eq:solution} allows the introduction of an extremely 
%\textcolor{red}{MT: I'm wondering that studying the model even without the feedback on the consumption propensity, i.e. a standard DSGE but with heterogeneities, might be  interesting. One could linearize around the unique equilibrium, compute future expectations, and be able to solve also the Euler's equation, which would provide an alternative DSGE framework compared to the TANKs. One could, for instance, study how the distributions of wages impact the economy.}

To move forward, we need to specify the distribution of skills $z^i$ over the population, as well as the dynamics of the overall productivity factor $\mathfrak{z}_t$.

Since in our model real wages are proportional to skills (see Eq.~\eqref{eq:wage}) we take inspiration from empirical data, which shows that wages follow an exponential distribution, except in the extreme tails where it becomes fatter (possibly Pareto-like), in part due to returns on investment, see e.g. \cite{tao2019exponential}. 
In order to keep the model as parsimonious as possible, we therefore assume that the distribution of $z^i$ in the population is given by:
\begin{equation}
\label{eq:zbar} 
\rho\left({z}^i\right) = 
\left \{
\begin{array}{ll}
     \frac{1}{\mu} \exp\left(\frac{{z}_{\min}-{z}^i}{\mu}\right) & z^i \geq z_{\min} \\
    0 & z^i < z_{\min}.
    \end{array}\right .
\end{equation}
This exponential distribution has a mean given by $\mathbb{E}[z]=z_{\min} + \mu$, which can be considered as a proxy for  the GDP per household of the corresponding economy. The distribution of wages is also characterised by a Gini coefficient~$\mathcal{G}$, which is a measure of the inequalities in our economy:\footnote{The Gini coefficient is defined as the average absolute difference between two randomly chosen individual, divided by the mean. By construction, the Gini coefficient is between 0 and 1.}
\begin{equation} \label{eq:gini}
    \mathcal{G} =  \frac{\mu}{2\mathbb{E}[z]} = \frac12 \, \frac{\mu}{\mu + z_{\min}}.
\end{equation}
Hence, $\mathcal{G} \to 0$ when $\mu \ll z_{\min}$ (egalitarian society) and $\mathcal{G} \to 50\%$ when $\mu \gg z_{\min}$. Stronger inequalities (i.e. $50\% < \mathcal{G} \leq  1$)  would require a different functional form, with, for example, extra power-law tails, or a Dirac mass at $z=z_{\min}$. The quasi-totality of European countries have a Gini index ranging between $24\%$ and $35\%$ while more unequal societies, such as the US, have Gini's $> 40\%$ \cite{ourworld2016gini}.  

Exogenous shocks are encoded into the idiosyncratic noise  $\mathfrak{z}_t$ that we write as $\mathfrak{z}_t = e^{\xi_t}$, where $\xi_t$  follows an AR(1) process:
\begin{equation} \label{eq:noise}
    \xi_t = \eta \xi_{t-1} + \sqrt{1-\eta^2}\mathcal{N}(0,\sigma^2) \ ,
\end{equation}
where we fix $\eta = 0.2$ (the parameter $\eta$ only affect the time-scale of the memory kernel of the stochastic process).
This corresponds to assuming that all individual productiveness $z^i$ are subject to the same exogenous shock. One could consider a richer model where different skills are affected by different shocks, but we leave such an extension for future investigations.

Note that the most probable value of $\mathfrak{z}_t$ is unity (i.e. $\xi_t=0$), which corresponds to what we will call ``normal'' or ``baseline'' conditions.

\section{Social Network and Self-Reflexivity}\label{sec:feedback}

We now discuss the specific form of the consumption propensity $f_t^i$ or, equivalently, its (re-scaled) square-root $F_t^i$ defined in Eq.~\eqref{def:F}. Following Ref.~\cite{guglielmo2019confidence}, we assume that the consumption propensity of agent $i$ at time $t$ depends on the realised consumption at time $t-1$  of some of the other agents of the economy (self-reflexivity), which are coupled to $i$ via an interaction network $J_{ij}$: 
\begin{eqnarray}
F_t^i \equiv \mathcal{F}^i\left( \frac{1}{K_i}\sum_{j (\neq i)} J_{ij} c_{t-1}^j \right), \quad K_i:=\sum_{j (\neq i)} J_{ij},
\end{eqnarray}
where $\mathcal{F}$ is a certain function the argument of which is the local average of the  consumption at time $t-1$ of ``neighbours'' on the network, i.e. the agents $j$ for which $J_{ij}$ is different from zero. The specific choice of this interaction network will be discussed in details below.  Here we start by focusing on the properties of the feedback function $\mathcal{F}^i$. In our recent work \cite{guglielmo2019confidence} we have shown that a generic S-shaped function 
suffices to induce multiple equilibria, with stochastic switches (corresponding to economic crises and recoveries) between them. As in \cite{guglielmo2019confidence} we choose a logistic function of the form:
\begin{equation} 
\mathcal{F}^i(x) = \frac12 \big[(\nu_{\max}^i-\nu_{\min}^i) \tanh\left(\theta^i(x-c_0^i)\right)  + (\nu_{\min}^i+\nu_{\max}^i)\big].
\end{equation}
The parameters $\nu^i_{\min} > 0$ and $\nu^i_{\max} > \nu^i_{\min}$ represent the minimum and maximum levels of labour that household $i$ can possibly provide (Eq. \eqref{def:F}); $c_0^i$ is a ``confidence threshold'' where the concavity of $\mathcal{F}(c)$ changes. Intuitively, $c^i > c^i_0$ tends to induce a high confidence state, while $c^i < c^i_0$ a low confidence state. $\theta^i > 0$ is the steepness of the function $\mathcal{F}^i$ close to the threshold level, setting the width over which the transition from low confidence to high confidence takes place, and is related to the agents' sensitivity to consumption's changes. 
%Here we set  $\theta^i=5,\ \forall i$.

In order to fully specify the model, we still need to define the interaction network %a generative model for the social network, 
i.e. the link variables $J_{ij}$. We base our choice on a number of studies indicating that households sharing the same level of wealth tend to cluster together (see for example \cite{bischoff2014residential, taylor2012rise}). For example, in large cities, the real estate market is such that people sharing a comparable level of income populate the same neighbourhoods, and therefore attend the same schools, facilities and many other public infrastructures. Following this reasoning, we %choose to 
set $J_{ij}=1$ with %some 
probability $p_{ij}$ and $J_{ij}=0$ with probability $1-p_{ij}$, %otherwise, 
where $p_{ij}$ is given by 
\begin{equation}\label{eq:probability}
    p_{ij} \propto \frac{C}{N} \exp\left({-\frac{\vert z^i - z^j \vert}{\lambda(z^i + z^j)}} \right)  .
\end{equation}
This implies that % , depending on the parameter  $\lambda$,
agents with similar wages (i.e. $|z^i - z^j| / (z^i + z^j)$ small) are more likely to be in contact (i.e. $J_{ij}=1$) %interact 
than agents in different social classes (i.e. $|z^i - z^j| / (z^i + z^j)$ large). The stratification and the level of segregation of the society is tuned by the parameter $\lambda$: when $\lambda \gg 1$, wage differences become irrelevant whereas when $\lambda \ll 1$ interactions are almost exclusively within the same social group. 
The factor $C/N$ in Eq.~(\ref{eq:probability}) ensures that each household interacts with a small average number $C$ of other households. In fact, to be more precise, in the following we will consider ``random-regular graphs'' of fixed connectivity $C$~\cite{wormald1999models}, which are defined as a graph chosen uniformly at random among all possible graphs of $N$ nodes such that each node has {\it exactly}  $C$ edges connecting it to its neighbours.

The procedure that we implement to build the network goes as follows: 
\begin{enumerate}
    \item We first assign a wage level $z^i$ to each of the $N$ nodes of the network, which are iid variables extracted from the distribution~(\ref{eq:zbar}).
    \item We build a random-regular graph of fixed connectivity $C$.%(see e.g.~\cite{krauth2006statistical} for a concrete algorithmic procedure to generate it). 
    \item The links are then rewired through a Monte Carlo algorithm. In order to keep the connectivity fixed we proceed as follows:  We assign to any configuration an energy equal to $H = \sum_{\langle i,j \rangle} |z^i - z^j| / (z^i + z^j)$, where $\langle i,j \rangle$ designates pairs of neighbouring agents. From the randomly generated graph, we pick at random two links, say $i \to j$ and $k \to \ell$ and we swap the connections to $i\to k$ and $j \to \ell$. % $p$ and an effective ``temperature'' $\lambda$. 
    We compare the energies of the old configuration,  $H^{\textrm{old}}$ with that of the rewired configuration,  $H^{\textrm{new}}$. The new configuration is kept with probability $\min(1,e^{(H^{\textrm{old}}-H^{\textrm{new}})/\lambda})$. % and we reject the move otherwise. 
    This process is repeated until a stationary state is reached. It is possible to show rigorously that the probability distribution at equilibrium is given by Eq.~\eqref{eq:probability}.
\end{enumerate}

While one could have made a number of alternative choices to model heterogeneities in both income and social interactions, we believe the setting introduced above is general and simple enough, and contains the essential features that we want to account for.

It is worth adding that in this paper we only focus on {\it symmetric} interactions matrices (i.e. if $j$ influences $i$ then $i$ equally influences $j$) and leave the generalisation to directed networks to a future investigation.

%% set up
\section{Parameter specifications}\label{sec:parameters}

In this section we propose reasonable and parsimonious specifications for the different parameters defined in the previous section. %A numerical exploration of alternative choices shows that we indeed capture all the qualitative features of the model with the specifications detailed in the next paragraphs. 

\subsection{Wage distribution}

The exponential distribution of wages, Eq.\eqref{eq:zbar} has two parameters, governing the average wage and the Gini coefficient. In order to disentangle the two effects, we  first investigate the model with a fixed value of $\mathbb{E}[z]$, arbitrarily set to $2$, and vary $\mu$ in the interval $[0.2,1.8]$, corresponding to Gini coefficients (given in this case by $\mathcal{G}=\mu/4$) between $5\%$ and $45\%$.
As discussed above, the average productivity level $\mathbb{E}[z]$ represents the average income and is essentially proportional to the GDP per capita of one country. When comparing the predictions of our model to real world data, we will relax the constant salary mass hypothesis and impose it to be proportional to the GDP/capita. This extension is discussed in detail in the last section of this work. 

\subsection{Feedback function}

The feedback function is specified, for each agent $i$, by four parameters: $\nu_{\max}^i$, $\nu_{\min}^i$,  $c_{0}^i$ and $\theta^i$. %The makes $4N$ parameters in total which is clearly unreasonable. 
We assume that the minimum amount of labour provided by an agent $i$, $\nu_{\min}^i$, is the same for each agent and equal to zero. (For practical convenience we fix it to a very small value,  $\nu_{\min}^i=10^{-3}$.) Similarly, the maximum amount of labour  $\nu_{\max}^i$ can be set to 1, independently of $i$. Using Eq. \eqref{eq:solution}, this implies that consumption $c^i$ in booming times is proportional to income $z^i$, as expected. %On the other hand we will introduce some behavioural dynamics allowing for heterogeneities in the choice of 

The most important parameters of the feedback functions are %the threshold consumption 
$c_{0}^i$ and %the sensitivity to economic fluctuations  
$\theta^i$. We assume that these parameters only depend on the income of each agent as detailed below. In this  way we are able to reduce drastically the number of free parameters of the model.

\begin{figure}[h!]
    \centering
    \includegraphics[width = 0.7\columnwidth]{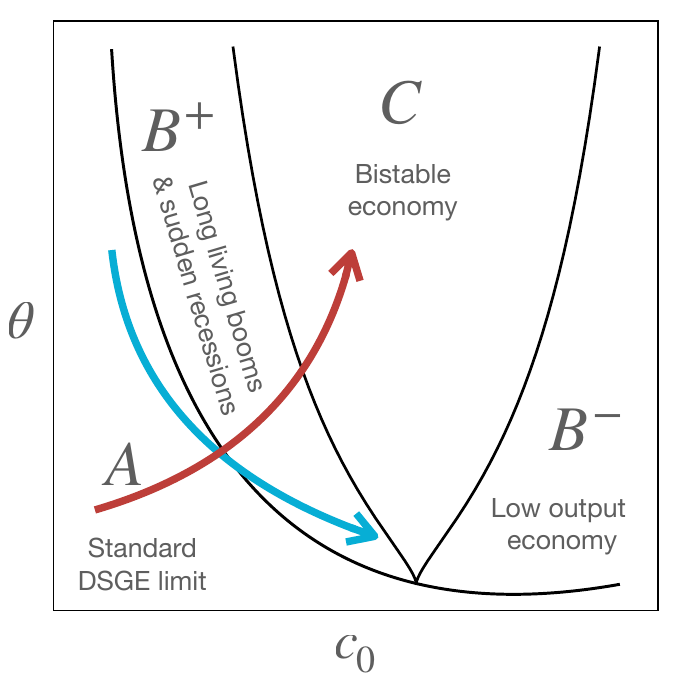}
    \caption{In the figure we show a sketch of the phase diagram as in the homogeneous case, with highlighted the different phases and their properties. In red and blue we draw two possibilities for the choice of the $\theta(c_0)$.  The arrows point in the same direction as the increase in wages' levels. This figure is meant to be a guide to help the reader to understand the model and the choice of  parameters.}
    \label{fig:fig1}
\end{figure}

The $c_0^i$'s play a key role as they correspond to the threshold %that represents 
below which the households' confidence collapses. In a recent article by D. Jacobe,\footnote{https://news.gallup.com/poll/111895/HigherIncome-Americans-Turning-More-Pessimistic.aspx} it is reported that in the early stages of the 2008 GFC the wealthier part of the population was also the most pessimistic about the state of the economy. To account for this effect, we set the confidence threshold of each household $c_0^i$ to be an increasing function of its income level, and hence of its productivity level $z^i$, modulated by an exponent~$\beta_1 \geq 0$: 
\begin{equation}\label{def:c0}
c_0^i = \bar{c}_0 \times (z^i)^{\beta_1} \, ,
\end{equation}
where $\bar{c}_0$ is a global trust level that 
we assume %that the global level of confidence $\bar{c}_0$ is 
to be determined by country specific economic policies, culture, etc. The larger the value of $\beta_1$, the stronger the dependence of the confidence threshold on income.

In order to gain some intuition about the specification of the sensitivity parameters $\theta^i$, we use as a guide the phase diagram established in the homogeneous case in~\cite{guglielmo2019confidence}, recalled in Fig.~\ref{fig:fig1}. Depending on the values  of $\theta$ and $c_0$ one can distinguish four zones in the phase diagram %Fig.~\ref{fig:fig1} 
that encode different properties of the economy. Phase $A$ delimits the area of the standard DSGE model, where we do not observe any economical crisis, zone $B^+$ allows for short-lived economical recessions, the $C$ phase admits a second equilibrium and correspondingly allows for crisis and economical recoveries with comparable probability and duration. Finally zone $B^-$ represents the set of parameters for which the system is systematically in a state of crisis. Over such a phase diagram we draw two possible ``trajectories'' for $\theta(c_0)$: in blue a convex decreasing relation and in red an increasing one. Both curves cross different phase transition lines, but the blue one seems a more natural choice. 
%as it corresponds to the wealthy class being less sensitive to the consumption of their social neighbours. % as it follows the phase diagram lines. 
Actually, we find that along the red curve it is almost impossible to find a set of parameters for which all the agents belong to the same phase and, moreover, the richest part of the population is systematically exposed to the economic crisis regardless of the choice of parameters. 

For these reasons we discard the ``red'' option and parameterise $\theta^i(c^i_0)$ as:
\begin{eqnarray}\label{def:th}
\theta^i(c_0^i)  = \bar{\theta} \times  (z^i)^{-\beta_2} = \bar{\theta} \times \left(\frac{c_0^i}{\bar{c}_0}\right)^{-\frac{\beta_2}{\beta_1}} \, ,
\end{eqnarray}
where $\bar{\theta}$ represents the global sensitivity scale and the exponent $\beta_2 >0$ enforces a monotonic decreasing dependence between $\theta^i$ and incomes.
When $\beta_2=\beta_1$, the width $\theta^{-1}$ of the transition region scales as the consumption threshold $c_0$ itself. When $\beta_2 < \beta_1$ on the other had (as we will find empirically), this width increases slower that $c_0$, meaning that high incomes are (on a relative basis) more sensitive than low incomes to a drop of consumption of their neighbours.  %This corresponds to the blue curve of figure Fig.\ref{fig:fig1}.
%In this scenario the role of central banks is key as, through forward guidance and narratives, they can influence the global levels  $\bar{c}_0$ and $\bar{\theta}$ affecting the probability of an economical recession.

Visually, when $\bar{\theta}$ increases the blue line is globally shifted upwards, while if $\bar{c}_0$ increases it  is shifted to the right.  
When $\beta_1, \beta_2 \to 0$, $c_0^i = \bar{c}_0$ and $\theta^i = \bar{\theta}$,  behavioural heterogeneities are switched off and the model leads to a phenomenology very similar to the one reported in \cite{guglielmo2019confidence}.  
%This choice reduces the dimension of the parameter's space from $2N \to 2$. 
We are thus left at this stage with only four parameters: $\beta_1, \beta_2$, $\bar{c}_0,\bar{\theta}$. Although seemingly restrictive, this setting gives rise a  rich phenomenology that we are going to analyse in the next sections.

\section{Characterising Crises Typologies}
\label{sec:crises_dynamics}
%\section{building some intuition}
\label{sec:intuition}

\subsection{Numerical results}

\begin{figure*}[t!]
    \centering
    \includegraphics[width = \textwidth]{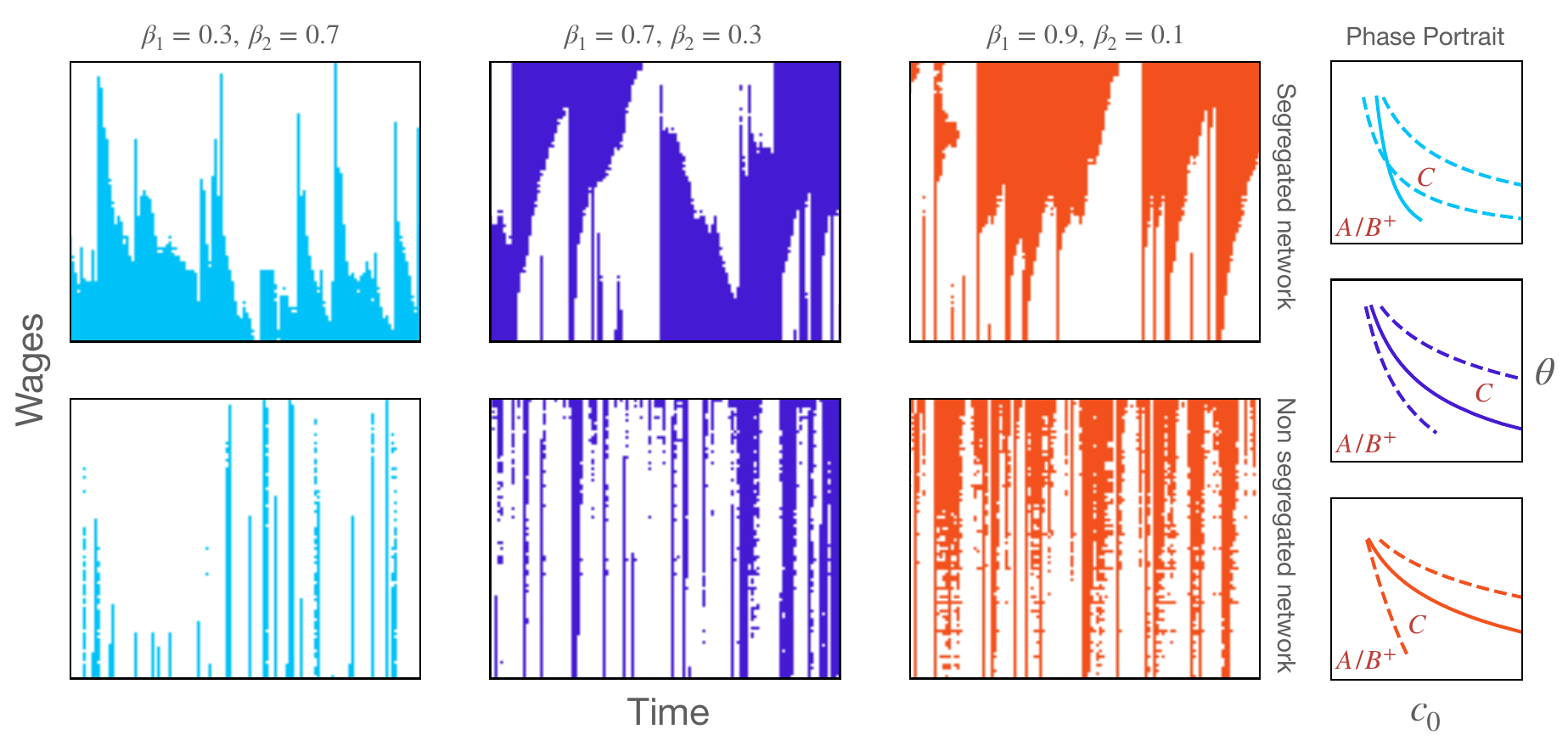}
    \caption{The graphs show the crises dynamics for three choices of the parameters, together with the relative phase diagram, using the same colour code. In the main panels, the abscissa is time and the ordinate are households, sorted by increasing wage. Colour appears when one agent's consumption drops below its corresponding threshold. In the 3 right panels sketching the phase diagram, the dashed line indicate the boundaries of the $C$ phase while the plain line represents the locus of $c_0$ and $\theta$ for different households. The three dynamics differ only by the choice of the couple $\beta_1$ and $\beta_2$, while the global levels of $\bar{\theta} = 4$ and $\bar{c}_0 = 0.5$ are kept constant, together with the level of income inequalities $\mu = 1.5$. For the upper set of graphs, the network is segregated, $\lambda=0.01$, while the bottom ones are with $\lambda = 100$. Note that the typology of crises changes substantially between the two cases.}
    \label{fig:fig2}
\end{figure*}

In our previous article \cite{guglielmo2019confidence} we showed how the introduction of the feedback function can destabilise the standard DSGE equilibrium. In the $C$ phase  the self-consistent solution  for the consumption has two fixed points allowing the system to switch from a high to a low consumption state. As shown in the phase diagram of Fig.~\ref{fig:fig1}, the confidence threshold $c_0$ modulates the probability of jumping from a high consumption state to a recession regime (this probability increases with $c_0$). In the present work this mechanism remains unchanged but the chain of events that bring the consumption of agent $i$ to collapse is more intricate. Figure \ref{fig:fig2} gives some insights about the possible scenarios. 

The three pairs of panels shown in  Fig.~\ref{fig:fig2} display the crisis propagation for three different choices of the parameters $\beta_1$ and $\beta_2$, for fixed values of $\bar{\theta}$ and $\bar{c}_0$. We also compare segregated (top row) and non-segregated networks (bottom row). In the segregated case we observe that crises form suddenly and then slowly abate.

Changing the values of $\beta_1$ and $\beta_2$, e.g. moving from the left to the right panels affects dramatically which social class undergoes consumption-driven crisis. In the left panels, for instance, crises spread almost exclusively from the poorest end of the population towards the middle class and only sporadically affect the whole system. In the central panels we observe a different scenario: the recessions originate with almost the same frequency from the richest or the poorest part of the population and affect both social classes with the same intensity.  Finally in the right panels crises always start from the richest agents and propagate towards the middle class and only in very few cases affect the whole population. 

In the bottom row of Fig.~\ref{fig:fig2} (non-segregated networks) we observe that
for the three choices of $\beta_1$ and $\beta_2$, only the shape and duration of the recessions are affected compared to the segregated case. Recessions spread more uniformly and are shorter.

Below we explain how to rationalise these observations.
%by analysing how heterogeneities deform the phase diagram shown in Fig.~\ref{fig:fig1}.

\subsection{A path across the phase diagram}

In the limit of strongly segregated networks ($\lambda \ll 1$), the only connections are between agents that have very similar income and, therefore, very similar values of $\theta$ and $c_0$. Their consumption obey a self-consistent equation very similar to the one studied in our previous paper \cite{guglielmo2019confidence}, but with wage dependent parameters:
\begin{equation}\label{eq:normal}
 c(z) = \gamma z\times  \mathcal{F}\left(c(z) \vert \theta(z), c_0(z)\right), \qquad \gamma := \frac{\mathbb{E}[z F]^{-\alpha}}{1-\alpha}.
\end{equation}
where the time dependence of the consumption $c$ is neglected in the absence of productivity shocks. 
As discussed in \cite{guglielmo2019confidence}, depending on the choice of the parameters $\theta$ and $c_0$, Eq.~\eqref{eq:normal} can have 1, 2 or 3 solutions. In the homogeneous case, the representative agent occupies a single point in the phase diagram of Fig.~\ref{fig:fig1}. In the limit of strong segregation, each social class occupies a different spot of the phase diagram. The union of these spots form the lines drawn in Fig.~\ref{fig:fig1}. %precisely describe this situation.  

The shape and the location of these lines strongly depends on the values of $\beta_1$ and $\beta_2$, as shown on the rightmost panels of Fig. \ref{fig:fig2} (the colour of each of those panels are chosen match the one of the corresponding dynamics). The top graph shows that agents with lower income are living in the $C$ phase (bi-stable economy) while the steep decrease of $\theta$ with $z$, allows the richer households to cross the $C \to B^+$ phase line. This reflects the dynamics shown on the corresponding panels.

In the middle panel the whole population lies within the $C$ phase, explaining why crises form from both sides of the income spectrum. Finally the bottom panel, although very similar to the previous case, reveals an important difference: households with a lower salary are closer to the line separating the phases $C$ and $A/B^+$. This affects the probability for the lower income class to suffer a drop of consumption, which is decreased compared to the middle panel case.

\subsection{The myopic effect of segregation}

The arguments given above are rigorous in the limit of segregated societies, but cannot explain the strong influence of $\lambda$ on the typology of crises. 
%. To explain those differences we have to analyse more carefully the role of $\lambda $ in the crises dynamics.
%The segregation level has a strong effect on the dynamics of the model, 
As %it is 
revealed by Fig. \ref{fig:fig2}, changing the structure of the interactions %network 
leads to %we have 
a drastic modification of the shape and duration of the recession spikes. %change. 
In fact, by varying the segregation of the network, we affect the correlation between the average income of the households (on which agents' trust  is based) and one's own income.

In a clustered society ($\lambda \ll 1$), the aggregate consumption of a family's neighbours is similar to the consumption of the agent itself. This creates an effect of myopia as agents probe the health of the economy only to a local scale. In this case, contagion effects are maximised. 
%If, therefore, a family's consumption suddenly collapses, its neighbours will have a drop in their feedback proportional to their own threshold.
As social segregation increases the fragility of the social class most exposed to an economical recession, as each agent is connected to others sharing a comparable wage and living in the same phase. 
%In other words this effect amplifies the role of the feedback which can produce deep and long crises even after a mild exogenous shocks. 
Hence we expect a sort of avalanche effect, %in the propagation of crises, 
as one agent's drop in consumption induces, with higher probability, the trust collapse of its neighbours. %\textcolor{red}{explain what ``jumps'' you're talking about. be more precise.}

On the contrary, in a non-segregated society agents base their trust in the economy by picking few agents chosen at random. This allows, for instance, the consumption of a low-wage person to be boosted by that of a wealthier neighbour, improving his own trust in the economy, and \textit{vice versa}. Diversification improves stability in this case: the domino effect is much weaker in the non-segregated network due to the fact that heterogeneous income level of the neighbours decreases the effects of the feedback function. Therefore the crises in the non-segregated case are shorter and rarer, and can only be produced by a stronger exogenous shock. 

The main message is thus that 
diversification of information sources increases resilience. 
In fact, comparing the top and bottom panels of Fig.~\ref{fig:fig2} we see that several small spikes in a non-segregated society coalesce in a unique recession event when segregation is strong, due to the avalanche effect described above. %\red{pour illustrer ce que tu dis là ça aurait pu être pas mal de mettre la même réalisation du bruit en haut et en bas dans la fig2.}

\subsection{Exogenous shocks and global crises}

\begin{figure}[t!]
    \centering
    \includegraphics[width = .49\textwidth]{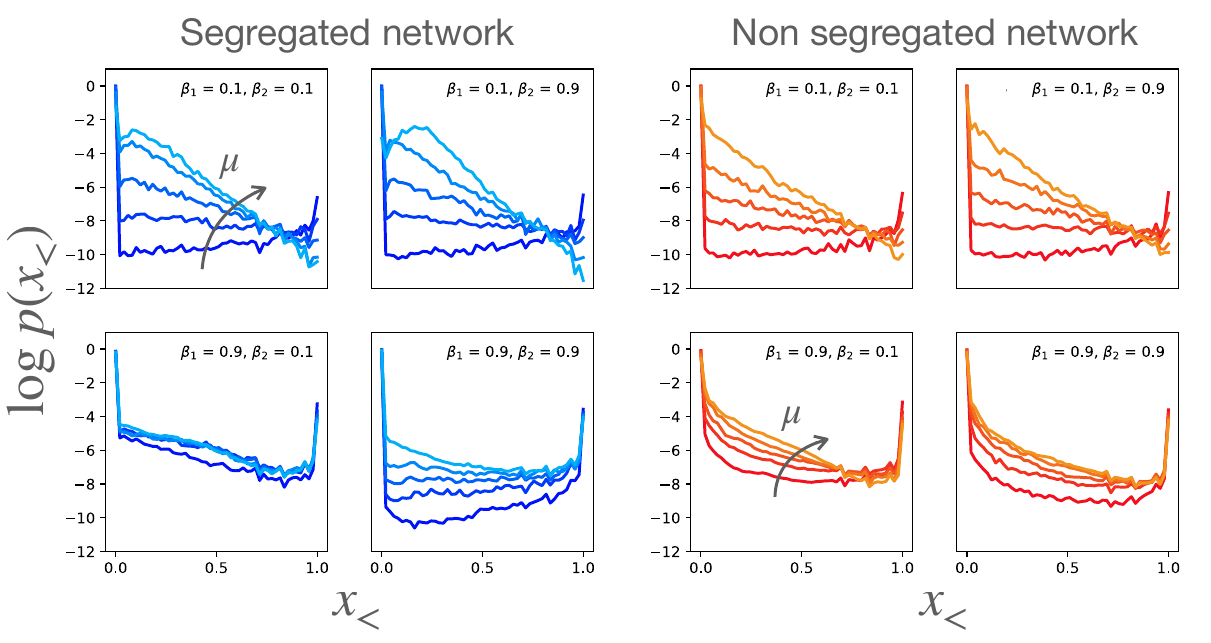}
    \caption{The panels show the probability distribution of the crisis size, $x_<$, for different choices of the parameters. The blue curves show the results for the segregated case, $\lambda = 0.01$, while the red ones represent the non-segregated scenario, $\lambda = 100$. Each panel is dedicated to a couple $\beta_1$, $\beta_2$ for five values of $\mu$ ranging from $0.2$ to $1.8$. $\bar{\theta} = 4$ and $\bar{c}_0$ are kept fixed.}
    \label{fig:fig4}
\end{figure}

After having investigated which households are the most affected by an economical recession, we now discuss how the size of the crises depends on those parameters, regardless of the social class. 
%In this paragraph we  discuss the crises sizes distribution as a function of the choice of the parameters. 
In order to do that, we introduce the quantity $x_{<,t}$, defined as the fraction of households being in a low consumption state at time $t$, independently of the income level:
\begin{equation}
    x_{<,t} := \frac1{N}\sum_{i=0}^N \Theta(c_0^i - c_t^i)\ ,
\end{equation}
where $\Theta$ is the Heaviside function: $\Theta(x>0)=1$ and $\Theta(x\leq 0)=0$. 
In the top panels of Fig.~\ref{fig:fig4} we draw the (logarithm of the) probability $p(x_<)$ of observing a crisis of ``size'' $x_<$, for different values of the income inequalities, of the exponents $\beta_1$ and $\beta_2$, and for segregated and non-segregated networks.

In the panels where $\beta_1 = 0.1$ (topmost panels) we observe a transition from a uni-modal to a bi-modal distribution as $\mu$ is decreased, i.e. as inequalities decrease. For low values of $\mu$ the probability distribution has two peaks: the first one in $x_< \approx 0$, describing a well functioning economy where most of the agents are in the high-consumption state, and the second one in $x_< \approx 1$, corresponding to global crisis where nearly all agents are in a recession state. In the uni-modal regime at larger $\mu$, instead, the probability becomes roughly exponential in the crisis size $x_<$.

%The crossover from the uni-modal to the bi-modal distributions is in fact associated to the presence (or the lack thereof) of  crises of intermediate size, that do not affect the whole population.

In the uni-modal regime most crises only affect a limited fraction of the population, and only very rarely hit the whole population (as in the examples shown in Fig.~\ref{fig:fig2}). Conversely, in the bi-modal regime recessions are mostly global.
This can be rationalised by recalling that in the limit $\mu \to 0$  all the agents have the same income, skills, and baseline consumption levels. We thus recover the results of the homogeneous model \cite{guglielmo2019confidence} in which only two states are possible (the whole population is in the good state or in the low-consumption one) and $x_<$ is either $0$ or $1$.

Introducing wage inequalities allows  for the possibility of having intermediate crises, that only affect a finite portion of the agents, thereby reducing the probability of a global crunch. 
 %\textcolor{red}{This might be explained thinking that the increase of the income inequalities prolongs the phase diagram reinforcing the effects of the other parameters. Questa frase non l'ho proprio capita.}

Comparing the right and left panels, we notice that the level of segregation does not have a major influence on the shape of the distributions $p(x_<)$ (even though the crisis dynamics itself is strongly affected by $\lambda$, as shown in Fig.~\ref{fig:fig2}).

At this point, the question that we still need to address is the following: what parameters affect, and how, the probability of having a global consumption crisis? 

In order to answer this question we introduce the probability $\mathcal{P}$ of observing a global crisis, which is defined as an event in which the consumption of more than $80\%$  of the population drops below their own level $c_0^i$, i.e.
\begin{equation}\label{def:P}
\mathcal{P} := \int_{0.8}^{1} p(x_<) \, \textrm{d}x_<.
\end{equation}
$\mathcal{P}$ plays the role of an order parameter for the uni-modal/bi-modal transition described above, as it is very small in the uni-modal regime and takes appreciable values for bi-modal distributions.

In our previous work~\cite{guglielmo2019confidence} we have shown that in the homogeneous case the crisis probability strongly depends on the amplitude of the external shocks $\sigma$ and on the global confidence threshold $\bar{c}_0$.
In the lower panels of Fig.~\ref{fig:fig4} we  plot the dependence of $\mathcal{P}$ on $\sigma$ for different choices of the other parameters. %the amplitude of the external shock 
 % fixing the %and the set of 
%other parameters ($\mu$, $\bar{c}_0$, $\bar{\theta}$, $\beta_1$ and $\beta_2$). 
For the sake of clarity, in each panel  we keep three of the  four parameters $\bar{c}_0$, $\bar{\theta}$, $\beta_1$ and $\beta_2$ fixed, and  let one of them vary (as indicated in the legends). In each panel we also show different curves corresponding to several values of $\mu$.

We find that the probability of having a global crisis becomes non zero beyond a certain  critical amplitude of the noise,  $\sigma_c$. We further observe that   $\sigma_c$ decreases with increasing $\bar{c}_0$ and/or $\bar{\theta}$. This result agrees with the simple intuition that for lower global confidence, or, similarly, stronger global sensitivity, global crises can be triggered by a smaller exogenous shock. In other words, referring again to the phase diagram shown in Fig.~\ref{fig:fig1},  an increase in $\bar{c}_0$ at constant $\bar{\theta}$ shifts the the system to the right, whereas an increase of  $\bar{\theta}$ shifts the system upwards. Households are thus pushed deeper into   the  $C$ phase and are more frequently exposed to global economic crises.
%that involve the whole population with higher frequency. 
\begin{figure}[t!]
    \centering
    \includegraphics[width = .49\textwidth]{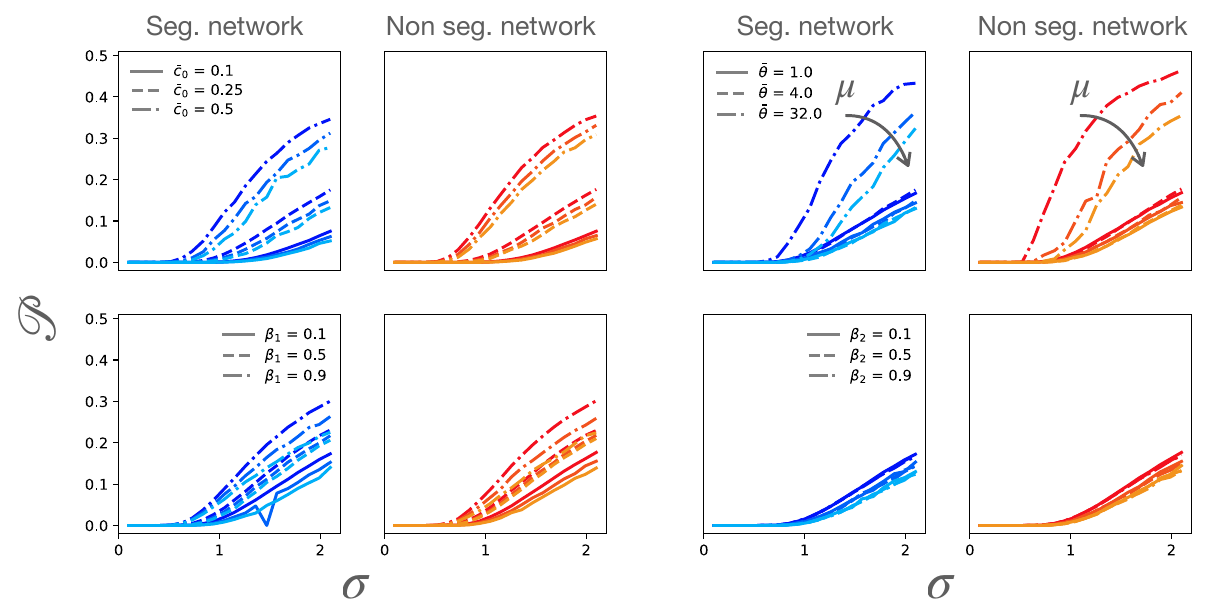}
    \caption{ The panels show the probability of a global crisis $\mathcal{P}$ as a function of the exogenous shock amplitude $\sigma \in [0.1,2.1]$. %In each panels the colour code reflect the income inequality $\mu$ as in Fig.~\ref{fig:fig4}. 
    We restrict the analysis to three values of $\mu$ : $0.2$, $1.0$ and $1.8$. In each panels the intensity of the colour reflects the income inequality $\mu$ as in Fig.~\ref{fig:fig4}. In each graph the global colour represents the level of segregation: when blue $\lambda = 0.01$, when red $\lambda  = 100$. We study the dependence of $\mathcal{P}$ on four parameters: in each couple of graphs (segregated and non segregated network) we let one parameter between $\bar{c}_0$, $\beta_1$, $\bar{\theta}$ and $\beta_2$ vary and we keep the other three constants. When kept constants the parameters take the following values: $\bar{c}_0 = 0.25$,  $\bar{\theta} = 4$ and $\beta_1 = \beta_2 = 0.1$. The varying parameter is represented with different line style, and the legend is shown within each panel.}
    \label{fig:fig4_bis}
\end{figure}
The effect of $\beta_1$ and $\beta_2$ on $\mathcal{P}$ and $\sigma_c$ is rather weak, as is the influence of segregation -- see Fig.~\ref{fig:fig4_bis}. Hence %This can be understood by recalling that  
$\beta_1$ and $\beta_2$ have an impact on the social class that is more frequently affected by the crises, but not on the probability of having a global recession. %which is, again, mostly controlled by the global levels $\bar{c}_0$ and $\bar{\theta}$.
%In fact $\beta_2$ modulates the response of the sensitiveness of the agents to its wage, steepening the . 

Finally, the same figure shows that increasing inequalities (higher values of $\mu$) generally lowers the probability $\mathcal{P}$ of having a global crisis, and increases the critical value $\sigma_c$. This is consistent with the content of the upper panels: as discussed above, increasing $\mu$ favours (in our model) the formations of smaller crises, that only affect a certain fraction of the population and reduces the exposure to global crises of the whole population.

%The order parameter $\mathcal{P}$ show a consistent dependence on then income inequality: in the panel showing the variation of $\mathcal{P}$ as a function of $\sigma$ for different values of $\bar{c}_0$, the increasing of $\mu$ also raise the critical value $\sigma_c$. This result is consistent with the upper panels of figure Fig.\ref{fig:fig3} where the behaviour of the probability distribution strongly depends on the inequality levels. More generally the value of $\mu$ influences the probability of having a global recession: the higher the inequalities the lower the cumulative probability $\mathcal{P}$ is. 

However the conclusion that more inequalities lead to a smaller probability of global crises is possibly misleading, as it neglects an important effect not accounted for in our model, namely the dependence of the global ``panic'' level $\bar{c}_0$ on the Gini coefficient $\mathcal{G}$. Indeed, a recent report from the OECD pointed out that ``\textit{societies with a strong middle class also experience higher levels of social trust (...). Today, however, middle-class households became increasingly anxious about their economic situation (...) given that middle incomes have not benefited from economic growth as much as upper incomes (...)}"~\cite{oecd2019middleclass}. In other words, higher income inequality should also raise the value of confidence threshold $\bar{c}_0$, leading to a more unstable society. It is not obvious which one of the two effects (stabilizing vs. destabilizing) is dominating. We in fact suspect that the influence of inequalities on 
$\bar{c}_0$ is non-linear, and only mild when inequalities are moderate.

\section{Empirical Data}\label{sec:data}

The results of the previous sections show that the model can reproduce a broad spectrum of possible scenarios for the formation and the propagation of economic crises across a society with stratified income levels.

In this final section we will exploit such versatility to compare the output of the model with real data, discussing differences and similarities when key parameters are modified. This exercise is not easy, as empirical data on the level of consumption for different income groups is not always available and/or complete for each country. On the other hand, data on income distribution exists.
Our aim here will be to exploit the available data to show that there is a region of the parameter space that is consistent with empirical observations on the relative drop of consumption of poorest compared to that of the richest during a crisis. 

However, since income data also includes returns from financial investments, our assumption that income has an exponential distribution is not adapted to describe the high tail, for which a power-law is more adequate \cite{tao2019exponential}. It may in fact be that a substantial part of the effect reported below results from financial losses, and not from the contagion effect captured by our model -- except perhaps in an effective way, see below. 

%\subsection{Observations and calibration}

The data  set we have explored is available from the website ``Our World in Data" \cite{owid}. It provides information regarding the consumption of the richest and the poorest decile, called respectively $c^a_{90,t}$ and $c^a_{10,t}$, for a large range of years $t$ and countries $a$. Furthermore, we refer to levels of GDP per capita, which is available on the same platform. The main interest of our study is to understand how heterogeneities and income inequalities affect the response of the population in a crisis scenario. 

The quantities $c^a_{90,t}$ and  $c^a_{10,t}$ are typically provided for each year $t$, but in cases where they are omitted, we interpolate the missing data point of the two closest available data points.\footnote{We use the logarithm for interpolation because we want to keep track of the exponential growth of consumption. For example, if the natural progression is $2,\ \star,\ 8$, where $\star$ represents the missing information, using this method we find $\star=4$, which seems more reasonable. Without interpolating the data, the number of points with complete information for GDP/capita, $\mathcal{G}$ and $\Delta$ is $113$. However, if we interpolate the missing information this number rises to $206$. $10$ of these countries have $\mathcal{G}$ greater than $50\%$ and are therefore not exploited, as our exponential model does not account for Gini's larger than $50\%$.}  
In order to track and compare the time evolution of the consumption of the highest and the lowest deciles, we compute, for each year $t$ and country $a$, the relative difference:
\begin{equation}
    \delta c^a_{\star,t} := \frac{c^a_{\star,t}-c^a_{\star,t-1}}{c^a_{\star,t-1}} \, , \quad \star =10,90  \ .
\end{equation}
It is clear from the definition that when $\delta c^a_{\star,t}$ assumes negative values it means that the consumption of the $\star$-th decile of country $a$ has dropped in the time lapse of one year. We define such an event as a recession that affected at least one extreme of the population, i.e. either $\delta c^a_{90,t} <0$ or $\delta c^a_{10,t} <0$. To monitor how unequally such crises affect the population, we introduce the indicator $\Delta_t^a$  defined as:
\begin{equation}
    \Delta^a_t = \delta c^a_{90,t} - \delta c^a_{10,t}  \, .
\end{equation}
This quantity $\Delta^a_t $ captures how economic crises spread in the society:
\begin{enumerate}
    \item If $\Delta^a_t < 0$ the richest decile undergoes a greater relative drop in consumption during the  crisis compared to the poorest decile.
    \item If $\Delta^a_t > 0 $ the poorest decile experiences the largest relative consumption drop.
\end{enumerate}

The other key elements of our model are the segregation index $\lambda$ (for which we have no direct data) and income inequalities, described by the Gini index $\mathcal{G}^a_t$ associated to country $a$ at date $t$. We also cut our sample into rich countries, with GDP/cap. larger than the median, and poor countries, with GDP/cap. less than the median.

%and the $GDP^a_t$/pers of that particular country measured at the year $t$.
%\red{ce serait bien de dire quelque part combien de données il manque, car interpoler comme ça alors qu'on veut mesurer des discontinuités c'est un peu moyen s'il y a trop de données manquantes.}
%As we don't distinguish between the years we will drop the time dependence: $\mathcal{G}_t \to \mathcal{G}$ and $\Delta_t \to \Delta$. 

%Result (click "Generate" to refresh) Copy to clipboard
% \usepackage{multirow}
% \usepackage{multirow}
% \usepackage{graphicx}

\setlength{\tabcolsep}{1.2mm}
\renewcommand{\arraystretch}{2}
\begin{table}[t!]
\centering
\resizebox{.47\textwidth}{!}{%
\begin{tabular}{|cc|cc|cc|}
\hline
\multicolumn{1}{|l}{} & \multicolumn{1}{l|}{} & \multicolumn{2}{c|}{\textbf{Segregated network}} & \multicolumn{2}{c|}{\textbf{Non segregated network}} \\
\multicolumn{1}{|l|}{\textbf{$\beta_1$}} & \textbf{$\beta_2$} & \multicolumn{1}{c|}{\textbf{GDP/cap. \textless med.}} & \textbf{GDP/cap. $\geq$ med.} & \multicolumn{1}{c|}{\textbf{GDP/cap. \textless med.}} & \textbf{GDP/cap. $\geq$ med.} \\ \hline
\multirow{5}{*}{\textbf{0.1}} & \textbf{0.1} & -0.0004 & -0.0007 & 0.0004 & 0.0002 \\ \cline{2-2}
 & \textbf{0.3} & 0.0018 & 0.0019 & 0.0286 & -0.0004 \\ \cline{2-2}
 & \textbf{0.5} & 0.0135 & 0.0146 & 0.2001 & 0.0009 \\ \cline{2-2}
 & \textbf{0.7} & -0.0032 & 0.009 & 0.4824 & 0.0009 \\ \cline{2-2}
 & \textbf{0.9} & -0.0425 & 0.0274 & 0.6972 & 0.0141 \\ \cline{1-2}
\multirow{5}{*}{\textbf{0.3}} & \textbf{0.1} & {\ul{ \textit{-0.0017}}} & {\ul{ \textit{-0.0022}}} & 0.0002 & -0.0017 \\ \cline{2-2}
 & \textbf{0.3} & {\ul{ \textit{-0.0011}}} & {\ul{ \textit{-0.0011}}} & 0.0001 & 0.0004 \\ \cline{2-2}
 & \textbf{0.5} & 0.0039 & 0.0021 & 0.0195 & -0.0007 \\ \cline{2-2}
 & \textbf{0.7} & 0.016 & 0.0074 & 0.2613 & -0.0011 \\ \cline{2-2}
 & \textbf{0.9} & 0.0367 & 0.036 & 0.4777 & -0.0002 \\ \cline{1-2}
\multirow{5}{*}{\textbf{0.5}} & \textbf{0.1} & -0.0006 & -0.0352 & 0.0 & -0.0959 \\ \cline{2-2}
 & \textbf{0.3} & {\ul{ \textit{-0.0021}}} & {\ul{ \textit{-0.0021}}} & 0.0006 & -0.004 \\ \cline{2-2}
 & \textbf{0.5} & {\ul{ \textit{-0.0008}}} & {\ul{ \textit{-0.0011}}} & 0.001 & 0.0012 \\ \cline{2-2}
 & \textbf{0.7} & 0.0192 & 0.0017 & 0.02 & -0.0008 \\ \cline{2-2}
 & \textbf{0.9} & 0.0394 & 0.0194 & 0.3325 & -0.0012 \\ \cline{1-2}
\multirow{5}{*}{\textbf{0.7}} & \textbf{0.1} & 0.0007 & -0.1153 & -0.0056 & -0.4378 \\ \cline{2-2}
 & \textbf{0.3} & 0.0004 & -0.0671 & -0.0019 & -0.1269 \\ \cline{2-2}
 & \textbf{0.5} & -0.0003 & -0.0045 & 0.0014 & -0.0067 \\ \cline{2-2}
 & \textbf{0.7} & -0.0007 & -0.0005 & 0.0015 & 0.0014 \\ \cline{2-2}
 & \textbf{0.9} & 0.0669 & 0.0023 & 0.0123 & -0.0002 \\ \cline{1-2}
\multirow{5}{*}{\textbf{0.9}} & \textbf{0.1} & -0.0065 & -0.0732 & -0.0172 & -0.6745 \\ \cline{2-2}
 & \textbf{0.3} & 0.0002 & -0.0674 & -0.0072 & -0.4255 \\ \cline{2-2}
 & \textbf{0.5} & 0.0015 & -0.043 & -0.0019 & -0.1203 \\ \cline{2-2}
 & \textbf{0.7} & 0.0011 & -0.005 & 0.0006 & -0.0124 \\ \cline{2-2}
 & \textbf{0.9} & 0.0009 & 0.0004 & 0.0015 & 0.0029 \\ \hline
\multicolumn{2}{|c|}{\textbf{Empirical data}} & \multicolumn{1}{c|}{-0.0018} & -0.0017 & \multicolumn{1}{c|}{-0.0018} & -0.0017 \\ \hline
\end{tabular}%
}
\caption{This table documents the coefficients of linear regressions of numerical $\Delta$s as a function of the Gini coefficient $\mathcal{G}$, for different choices of parameters: $\beta_1$, $\beta_2$  and  $\lambda=0.01$ (segregated) or $\lambda=100$ (non segregated). We fix as constants: $\bar{c}_0 = 0.5$, $\bar{\theta} = 4$, $\sigma = 1$.
    For each combination of parameters, several independent simulations are performed, during which the time evolution of $ \Delta_t$ is calculated and then averaged, conditioned to a crisis, i.e. either $\delta c_{90}<0$ or $\delta c_{10}<0$. We further distinguish between countries having a GDP/cap. higher and lower than the median value of the available data. The reference value of the regression for empirical data is displayed at the bottom row of the table.}
\label{tab:tab1}
\end{table}
 
\begin{figure}[t!]
    \centering
    \includegraphics[width = .49\textwidth]{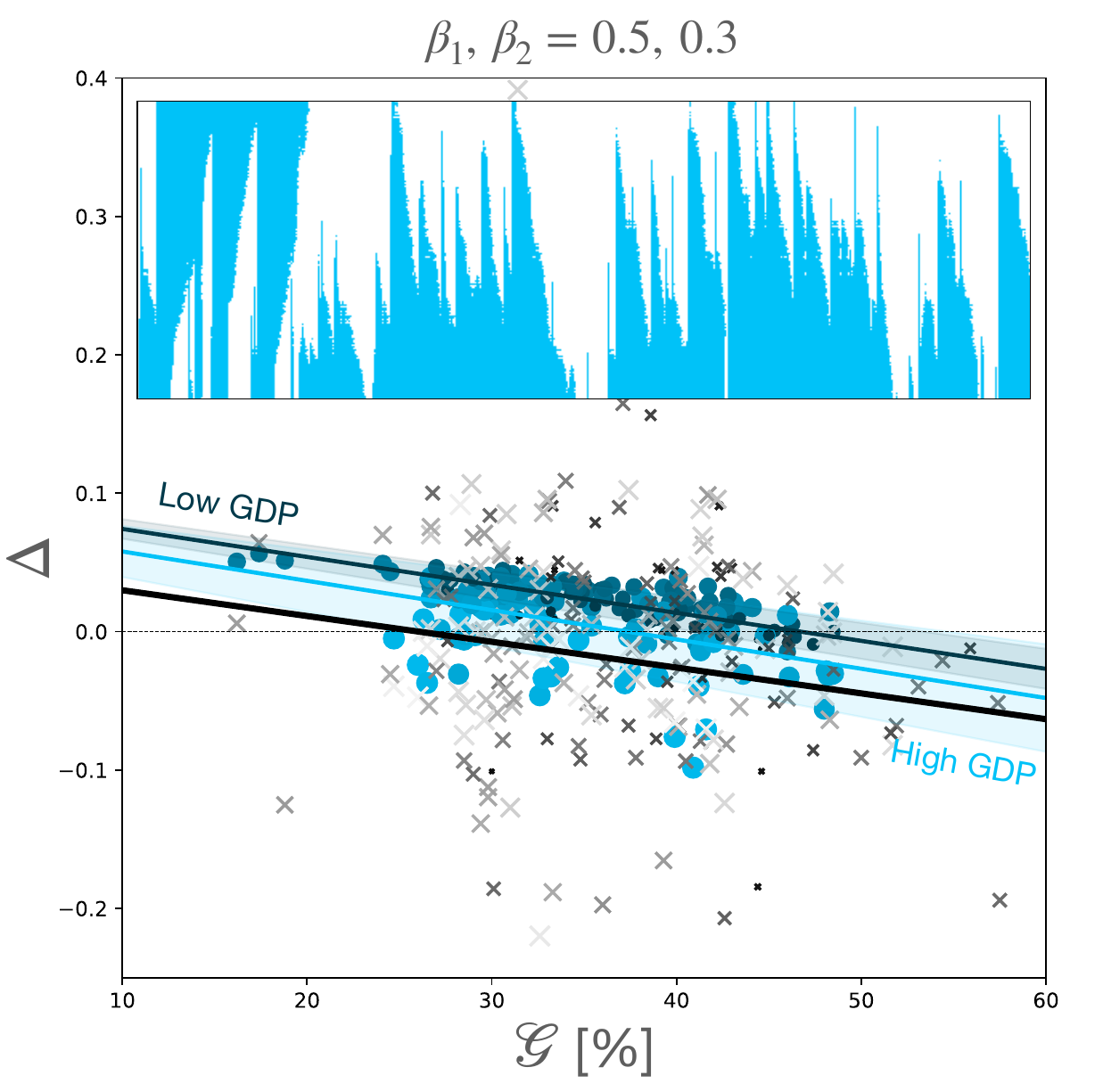}
    \caption{The panel shows the numerical simulation corresponding to the parameters that best fit real data, i.e. $\beta_1=0.5$, $\beta_2=0.3$, $\lambda=0.01$. The brightness and the size of each point is proportional to the GDP per capita: the bigger (or brighter) the dot is the stronger is the economy it represents. In each panel the markers (`x') correspond to real data and are shown in shades of grey. The solid black line is the linear regression through real data, which is found to be very similar for rich countries and for poor countries.
    %(for the sake of clarity we only show the regression for countries with GDP/pers $<$ median). 
    The horizontal dashed black line shows the reference line $\Delta=0$. The two coloured lines represent the linear regressions (errors on the regressions are also displayed as coloured bands) for low GDP/cap. countries (dark blue) and high GDP/cap. countries (light blue), again very similar to one another and to real data. The upper panel shows a numerical realisation of the crisis dynamics, as in Fig.~\ref{fig:fig2}, for the same values of $\beta_1, \beta_2$, $\lambda$ and with $\mathcal{G}=0.411$ and $\mathbb{E}[z]=10$, corresponding to the US economy in 2016 \cite{world_bank_gini}. In this example, low income households are more frequently in a low consumption state, although occasional crises also hit high earners (see left part of the time series). 
    %In the right inner panel we display the phase portrait ($\log-\log$ scale) correspondent to the best choice of parameters, light blue line, and two other possibilities: violet line is obtained reducing the term $\bar{c}_0$ and the red one by increasing it. 
    }
    \label{fig:fig5}
\end{figure}
The processed data is displayed in Fig.~\ref{fig:fig5} where we show $\Delta^a_t$ versus $\mathcal{G}^a_t$, for all available years $t$ and countries $a$ (without distinctions).
To better visualise the GDPs we set the size of the markers (x) proportional to its value and we choose to adapt the grey level accordingly: light grey corresponds to high GDPs, and dark grey to low GDPs. For the following discussion we will refer to $\Delta$ as being the set given by $\Delta_t^a,\forall a,t$.

We observe that $\Delta$ exhibits a {\it negative} overall correlation with $\mathcal{G}$: $\mathbb{C}(\Delta, \mathcal{G}) \approx -0.126$. This means, perhaps unexpectedly, that with the increase of inequalities the relative response to a recession is in favour of the poorest.\footnote{A double regression against both Gini and GDP/cap. shows that the direct impact of GDP on $\Delta$ can be safely neglected.}
This is compatible with our assumption that $\beta_1 > 0$, i.e. that the confidence threshold of the high earners is higher than that of the low earners (meaning that transition to a low consumption state is more probable for higher wages). Indeed, it is difficult for low incomes to reduce what is already the bare minimum consumption. 

In order to calibrate the model in a realistic way as to reproduce these observations, %and to be the more realistic possible, we calibrate our model as follows: every data point has associated a Gini index and the GDP/pers.
%we take up an idea already mentioned in the section \ref{sec:parameters}:
%data says that the average income (which, in turn, is proportional to the GDP per capita) correlates with inequalities \cite{causa2014can}; in fact, developing economies are often characterised by very high Gini  scores $\mathcal{G}$ and lower incomes compared to developed countries.\footnote{ There are exceptions to this trend, for example US have large Gini and large GDP/pers.} 
%When considering an exponential distribution of the wages (as we do), Eq. \eqref{eq:gini} encodes the dependence of the Gini index on the salary mass. We therefore 
we drop the fixed average salary $\mathbb{E}[z]$ hypothesis (which has been used in the previous sections to explore the possible scenarios of the model) and we %we perform each run with a fixing 
set $\mathbb{E}^a[z] \propto \, \text{GDP}^a$/capita.\footnote{Many developed countries have social policies that allow to reduce the confidence threshold via social aids of the welfare system that increase the global trust in the economy. Those policies can be modelled, for example, introducing a new parameter $\beta_3$ that modulates how $\bar{c}_0$ of a country scales with the  the GDP/capita, setting for example $\bar{c}_0 \to \widetilde{{c}}_0  \mathbb{E}[z]^{\beta_3}$, where $\widetilde{{c}}_0$ represents an arbitrary global confidence level. We have explored this extension of the model but systematically find that $\beta_3 \approx 0$ gives the best agreement with data.}
%The parameter $\beta_3$, when negative, is intended to account for policies that are mostly in effect in developed countries with higher GDPs.  They allow to reduce the confidence threshold via social aids of the welfare systems that increase the global trust in the economy.
 %On the opposite scenario, if simulations where $\beta_3 \geq 0$ are more realistic, we will conclude that higher the wealth of one country the faster its population react to an external shock. 
%On the contrary, $\beta_3 \geq 0$ corresponds to a global confidence threshold of a society which increases with the wealth of the respective country (the richer is the country the more ``souspicious'' are the individuals). 
%In the following we will investigate several values of $\beta_3$ in the interval $ [-0.5, 0.5]$.

The GDP/cap. of the United States will be used as a reference for the other countries. Without loss of generality we fix the average wage in the US to some arbitrary value, say $\mathbb{E}[z]^{\rm US} = 10$.
Having thus fixed the value of $\mathbb{E}^a[z]=10 \times \text{GDP}^a/\text{GDP}^{\rm US}$ and the Gini coefficient $\mathcal{G}^a$, the value of $\mu^a$ is uniquely determined by Eq. \eqref{eq:gini}.   

For definiteness, we set the global sensitivity $\bar{\theta} = 4$, the global confidence level $\bar{c}_0 =0.5$ and the amplitude of the noise to $\sigma=1$, independent of $a$. $\bar{\theta}$ and $\bar{c}_0$ can be changed quite a bit without affecting the quality of the final result, provided $\beta_1$ and $\beta_2$ are slightly modified as well. The value of $\sigma$ cannot be too low (otherwise crises almost never happen) nor too high (otherwise crises are too frequent), so $\sigma=1$ is a reasonable compromise. The most relevant parameters turn out to be the segregation level $\lambda$, and the exponents $\beta_1$ and $\beta_2$, which we scan but again uniformly across all countries. 

%\subsection{Results} 
We test our model for different combinations of the parameters running several simulations, each of which is based on the empirical data.
Unlike the real data, where GDP/capita does not influence much $\Delta$ (the linear regression has a coefficient of $-1.8 \times 10^{-3}$ when GDP/capita $<$ median and $-1.7 \times 10^{-3}$ otherwise), our simulations give a linear regression that depends quite strongly on GDP/capita.

We thus split our analysis of the correlations between $\Delta$ and ${\cal G}$ into countries having a GDP per capita greater and smaller than the median of the points considered. 
The results for the numerical values of the linear regressions of the outcome of our simulations are listed in Tab.~\ref{tab:tab1}, together with the parameters explored.  

%%HERE

We observe that the calibration of our model is very sensitive to the choice of $\beta_1$ and $\beta_2$, as the results differ greatly from case to case.
 Only for some values of the parameters do the simulations display a negative correlation between $\Delta$ and $\mathcal{G}$ independent of the level of GDP/capita. All other combinations of parameters are unrealistic and therefore discarded. We observe in particular that in the non-segregated scenario ($\lambda=100$) there is no choice of $\beta_1$, $\beta_2$ that is compatible -- even qualitatively -- with empirical values. 

On the other hand, when considering a segregated network (i.e. $\lambda = 0.01$), when $\beta_1 = 0.3$ and $\beta_2 \in \{0.1,0.3\}$, or $\beta_1 = 0.5$ and $\beta_2 \in \{0.3,0.5\}$ (underlined and in \textit{italics} in Tab.~\ref{tab:tab1}) our results are consistent with empirical data, both in terms of sign and magnitude. The results corresponding to the best case scenario is superimposed to real data in Fig.~\ref{fig:fig5}. (We set the brightness of each point dependent to GDP/capita: the greater the brighter.)

% We note, for instance, that when $\beta_1 \approx \beta_2$, the linear regressions coefficients are almost independent on the GDP/capita.

The role of segregation is quite an interesting outcome of our calibration exercise. It suggests, as is intuitively plausible, that contagion effects are mostly within social classes, and less across social classes. As we noted above, our model does not properly account for financial crises, which chiefly affects the high income class. However, a segregated network allows one to describe in an effective way the correlation in high income consumption shocks. 

%%%HERE
%These observations leave room for a brief discussion on how to perform a  better calibration of the  model. Although this is beyond the scope of this article and will  be left for future investigations,
%the differences that we observe between real data and the numerical simulations of the model
%This differences we found in the calibration according to the GDP per capita of countries 
%may be due to several factors that are  worth mentioning.

A better way to model these shocks would be to allow productivity shocks $\xi_t$ (defined in Eq.~\eqref{eq:noise}) to be correlated between individuals belonging to the same social class, with a variance also depending on outcome (and therefore on countries as well, through GDP/capita). We leave this for further investigations, as one would need more micro data to calibrate such an extended model. 

%The first issue of the calibration is that we  consider the same noise level $\sigma$  for countries having a different economical structure. To the contrary, it is reasonable to assume that  stronger economies also absorb exogenous shocks better, and consequently $\sigma$ should be re-adapted and calibrated as a function of the GDP per capita.

%The second issue is that probably the exponential distribution of wages
%, although representing well the poorest part of the population, 
%does not accurately reproduce the incomes of the richest individuals of the population (which should instead be better described as power-laws).
%, as it does not
%it is not accurate when reproducing the real wage distribution, and is not able to % as it 
%account for high revenues on the capital which are %and it is 
%strongly affected by the financial sector, which we are neglecting in this model. 
%To account for this effect we should exploit some ``\textit{ad hoc}'' shapes for the wage distribution that vary from country to country. 
%Those modifications can lead to a better calibration of this heterogeneous DSGE. Our intention is not to be predictive, but to allow DSGE models to take into account the heterogeneity present in the real world while giving consistent results that go in the same direction as real data. 
 
%To cite an example, an analysis conducted by New York times \cite{nytimes2020covid19} reports how, during the COVID-19 crisis, the richest quartile has faced an over 35\% drop in consumption, more than 5\% greater when compared to the poorest quartile, suggesting that $\Delta < 0$.

\section{summary, conclusion}

Let us summarise what we have achieved in this paper. First, we have extended the self-reflexive DSGE framework to heterogeneous households, which differ by their income level and by their social network. Consumption is therefore also heterogeneous and is given by Eq.~\eqref{eq:solution}, which appears to be new. Confidence feedback is mediated through the social network of each agent, which we assume to be either within social classes only (segregated network), or across social classes (non segregated network), with a parameter allowing one to smoothly interpolate between these two extremes. Depending on the specification of the confidence feedback function, we find a rich variety of possible crises types: propagating mostly within high income households, or mostly within low income households, or else, in a narrow parameter region, across the whole society. Interestingly, we find that crises are more severe for segregated networks, for which contagion effects are stronger. Inter social class interactions tend to blunt the propagation of pessimism, because agents belonging to different social classes have different sensitivities to economic shocks. We also find that more income inequalities lead to a small probability of global crises (all other parameters being kept fixed). However, this conclusion should be taken with a grain of salt, as other effect affecting confidence (such as insecurity, social violence, etc.) are not accounted for in the model -- although there is room to extend the model in that direction as well.

Finally, we have compared the prediction of the model with real data, that quantify the relative drop of consumption of the lowest income decile vs. highest income decile during recessions. Perhaps counter-intuitively, we find in more unequal countries (with high Gini coefficients), the consumption of the lowest income households tend to drop less than that of the highest incomes. This trend is mostly driven by the Gini coefficient and not by the country GDP/capita. Our model can be calibrated to reproduce such an empirical finding -- in fact only a small region of the parameters is compatible with the sign of the empirical effect. In particular, we find that the segregated network hypothesis is strongly favoured by the data, although other mechanisms, like financial market fluctuations, may lead to similar effects.

As with all models, many  possibly relevant features are left out in our model. We however hope that the framework proposed here -- which allows to mix together income inequalities and confidence feedback mediated by heterogeneous social networks -- can be welded with other approaches, such as popular HANK models for example \cite{kaplan2018monetary}. This would improve our understanding of economic recessions and their impact on the different strata of the society.  

%\footnote{https://www.ncbi.nlm.nih.gov/pmc/articles/PMC6218958/pdf/12889\_2018\_Article_6076.pdf}
\vskip 0.3cm

\paragraph*{Acknowledgements.}
We thank L\'eonard Bocquet, Ben Moll, Jos\'e Moran and Francesco Zamponi for many insightful discussions on these topics.
This research was conducted within the \emph{Econophysics \& Complex Systems Research Chair}, under the aegis of the Fondation du Risque, the Fondation de l'Ecole polytechnique, the Ecole polytechnique and Capital Fund Management.

\vskip .2cm\noindent
\bibliographystyle{unsrt}
\bibliography{biblio}

\end{document}